\documentstyle[aabib,psfig]{l-aa}
\include{epsf}

\newcommand{\msun}{\mbox{$M_\odot$}}

\newcommand{\rsun}{\mbox{$R_\odot$}}

\newcommand{\Rl}{Roche lobe~}

\begin{document}
\thesaurus{08.02.1;
           08.05.3;
           08.14.1;
           13.25.5}
\baselineskip 1.0 \baselineskip
\title{The formation of black-holes in low-mass X-ray binaries}
\author{Simon F.\ Portegies Zwart\/$^1$,   
        Frank Verbunt\/$^1$ \&\ 
        Ene Ergma\/$^2$}
\offprints{Simon Portegies Zwart}
\institute{$^1$ Astronomical Institute, Utrecht, Postbus 80000,
           3508 TA Utrecht, The Netherlands \\
           $^2$ Tartu University, Physics Department,
                {\~U}likooli 18 EE2400 Tartu, Estonia}     
\date{received; accepted:}
\maketitle
\markboth{S.F.\ Portegies Zwart, F. Verbunt \&\  E.\ Ergma}
         {Black-Holes in Low-Mass X-ray Binaries}

\begin{abstract}
We calculate the formation rates of low-mass X-ray binaries with
a black hole. Both a semi-analytic and a more detailed model
predict formation rates two orders of magnitude lower than 
derived from the observations. Solution of this conundrum 
requires either that stars with masses less than 20~\msun\ can
evolve into a black hole, or that stellar wind from a member of
a binary is accompanied by a much larger loss of angular momentum
than hitherto assumed.
\end{abstract}

\keywords{binaries: close -- stars: evolution -- stars: neutron -- 
          X-rays: stars}

\section{Introduction}
Six low-mass X-ray binaries currently are known to
have a mass function that indicates a mass of the compact object larger
than $\sim 3 \msun$, and these are therefore 
believed to have a black hole accretor.
All these systems are soft X-ray transients and
the total number in the galaxy of such systems is 
estimated to be between a few hundred and a
few thousand; at any moment in time only a small fraction is
X-ray active (for reviews, see \cite{cow92}, \cite{tl95}, \cite{ts96}).

The most popular formation scenario for low-mass X-ray binaries 
proposes a relatively wide binary 
with an extreme mass ratio as the progenitor system (\cite{vdh83}). 
The massive star evolves to fill its Roche lobe and engulfs its 
low-mass companion.
Due to the extreme mass ratio the low-mass star spirals into the envelope
of the high-mass star.
A close binary remains if the spiral-in ceases before
the low-mass companion coalesces with the compact helium core of the
primary. 
The helium core continues its evolution and may turn into
a neutron star or a black hole.
Only a few studies apply this evolutionary scenario specifically to the
formation of low-mass X-ray binaries with an accreting black hole
(see \cite{rom92}).

In this paper we discuss some problems of 
this standard formation scenario in its simplest form (Sect.~2) and
show that these persist in a more refined treatment (Sect.~3).
We quantify this in Sect.~4 and in Sect.~5 the results are discussed.

\section{Semi-analytic approach}
In the standard scenario the formation of a low-mass X-ray binary
starts with a high-mass primary and a low mass
secondary star ($m_o \sim 1 \msun$) in a detached binary.
For simplicity we assume that the initial orbit is circular.
As the primary evolves and expands to giant dimensions it 
fills its Roche lobe and starts to transfer mass onto its companion.
Due to the small mass ratio mass transfer is highly unstable and
results in a common-envelope phase.
A close binary remains provided that the primary's envelope is fully
ejected before the secondary star coalesces with the
compact core of the primary.
The reduction in the semi-major axis during the spiral-in can be computed
by comparing the binding energy of the primary's envelope with the
orbital binding-energy of the binary (\cite{web84}).
The helium core continues its evolution and finally collapses
into a neutron star or a black hole.
The binary becomes an X-ray binary once the secondary star fills its
Roche lobe.

We illustrate this scenario for primary stars of initially $M_o=20$ 
and 60~\msun which reach a maximum radius of
1000~\rsun (see \cite{rom92}).
The scenario requires that the primary fills its Roche lobe during its
evolution, i.e.\ that its Roche lobe has a radius less than 1000~\rsun.
The corresponding semi-major axis of the binary orbit may be calculated 
using the equation for the radius $R_L$ of the Roche lobe
given by \cite*{egg83}:
\begin{equation}
\frac{R_L}{a} \equiv {\cal R}_L(q) = {0.49\over 0.6+q^{2/3}\ln (1+q^{-1/3})}.
\label{Roche_lobe}\end{equation}
Here $a$ is the semi-major axis of the binary and $q \equiv m/M$ is the
mass ratio and $R_L$ is the Roche-lobe radius for the star with mass $M$.
For the 20 and 60~\msun primaries we thus find a semi-major axis of
about 1590 and 1440~\rsun, respectively.

The scenario further requires that the binary survives the 
spiral-in that follows upon the first Roche-lobe contact, 
i.e.\ that both the helium core and the 1~\msun companion star are
smaller than their Roche lobes.
The mass $M_{\rm c}$ and radius $R_{\rm c}$ of the helium core of the primary 
can be computed with (see \cite{it85} and \cite{dld92}, respectively)
\begin{equation}
M_{\rm c} = 0.073M_o^{1.42},
\label{Mcore}\end{equation}
and 
\begin{equation}
\log R_{\rm c} = -1.13 + 2.26\log M_{\rm c} - 0.78 (\log M_{\rm c})^2.
\label{Rcore}\end{equation}
The mass and radius of the secondary star are not affected by the spiral-in.
The mass ratio after the spiral-in gives the sizes of the Roche lobes
of the helium core and the secondary in units of the semi-major axis. 
From the radii of the helium star and its companion we can thus derive
the minimum semi-major axis for which both stars fit inside their
Roche lobes.
For the 20 and 60~\msun primaries we thus find a semi-major axis after
the spiral-in which should exceed 4.0 and 6.3~\rsun, respectively.
In both cases the most stringent limit is set by the main-sequence star.

The ratio of the semi-major axes before and after the spiral-in
can be computed by comparing the binding energy of the primary's envelope
with the binding energy released by the shrinking binary 
(see \cite{web84}):
\begin{equation}
{a_f\over a_i} = {M_{\rm c}\over M} \left( 1+ {2a_i\over\alpha\lambda R}
               {M_{\rm e}\over m} \right)^{-1}.
\label{spiral_in}\end{equation}
Here $a_i$ and $a_f$ are the semi-major axes at the onset and end of
the spiral-in, and $M$ and $R$ are the mass and radius of the
primary at the onset of spiral-in.
We set $\alpha\lambda=0.5$.
We assume that the primary did not lose any mass before it fills its 
Roche lobe: $M=M_o$ and $M_{\rm e} = M_o - M_{\rm c}$.
For the 20 and 60~\msun primaries we thus find a semi-major axis before
the spiral-in which should exceed 1470 and 3160~\rsun, respectively.

For the 20~\msun primary, the lower limit on the semi-major axis for
survival of the spiral-in is smaller than the upper
limit for Roche-lobe contact and a low-mass X-ray binary can be formed when
the initial semi-major axis is between these two limits.
For the 60~\msun primary, the lower limit on the semi-major axis for
survival of the spiral-in is larger than the upper
limit for Roche-lobe contact and no initial orbit leads to the formation
of a low-mass X-ray binary: Roche-lobe overflow always leads to a merger.

\begin{figure}
\hspace*{0.5cm}
\epsfxsize = 4.0cm
\epsffile{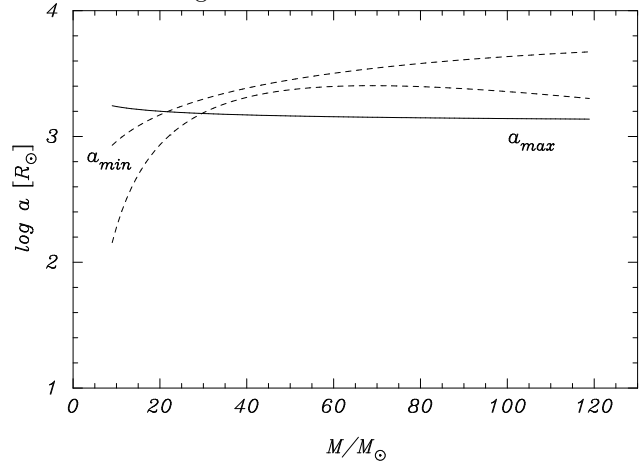}
\caption[]{Lower limit to the initial semi-major axis at which the
binary survives the spiral-in, as function of the mass of the primary. 
The upper (lower) dashed line gives the limit determined from the
condition that the secondary star (helium core) is smaller than its 
Roche-lobe, 
The solid line gives  the upper limit at which the primary 
reaches its Roche lobe at its maximum radius of 
$R_{max} = 1000\rsun$. The secondary is assumed to be a 1~\msun star}
\label{Fig_rom92}\end{figure}

Figure~\ref{Fig_rom92} shows, as function of
the mass of the primary, the lower and upper limits 
to the semi-major axis of the initial binary
at which the binary survives the spiral-in and the primary reaches the
Roche lobe.
Only for low mass primaries ($M \la 22\msun$)
does the binary survive the spiral-in.
Thus, if only stars with an initial mass larger than 40~\msun (\cite{hh84})
form a black hole, the formation of a low-mass X-ray binary with a black
hole as a compact object is excluded in the standard scenario.

One may argue that the binary survives the spiral-in even if the
main-sequence star is larger than its Roche lobe after spiral-in, 
as long as the helium star fits inside its \Rl.
The main-sequence star can shrink within its Roche lobe by transferring mass 
to the helium core. This mass transfer is stable because the
main-sequence star is less massive than the helium core.
In Fig.~\ref{Fig_rom92} we also show the lower limit to the semi-major
axis obtained from the condition that only the helium star fits inside its 
Roche lobe.
We see that even in this case low-mass X-ray binaries with a black
hole cannot be formed in the standard scenario.

\section{Detailed model}
\begin{figure}
\hspace*{0.5cm}
\epsfxsize = 4.0cm
\epsffile{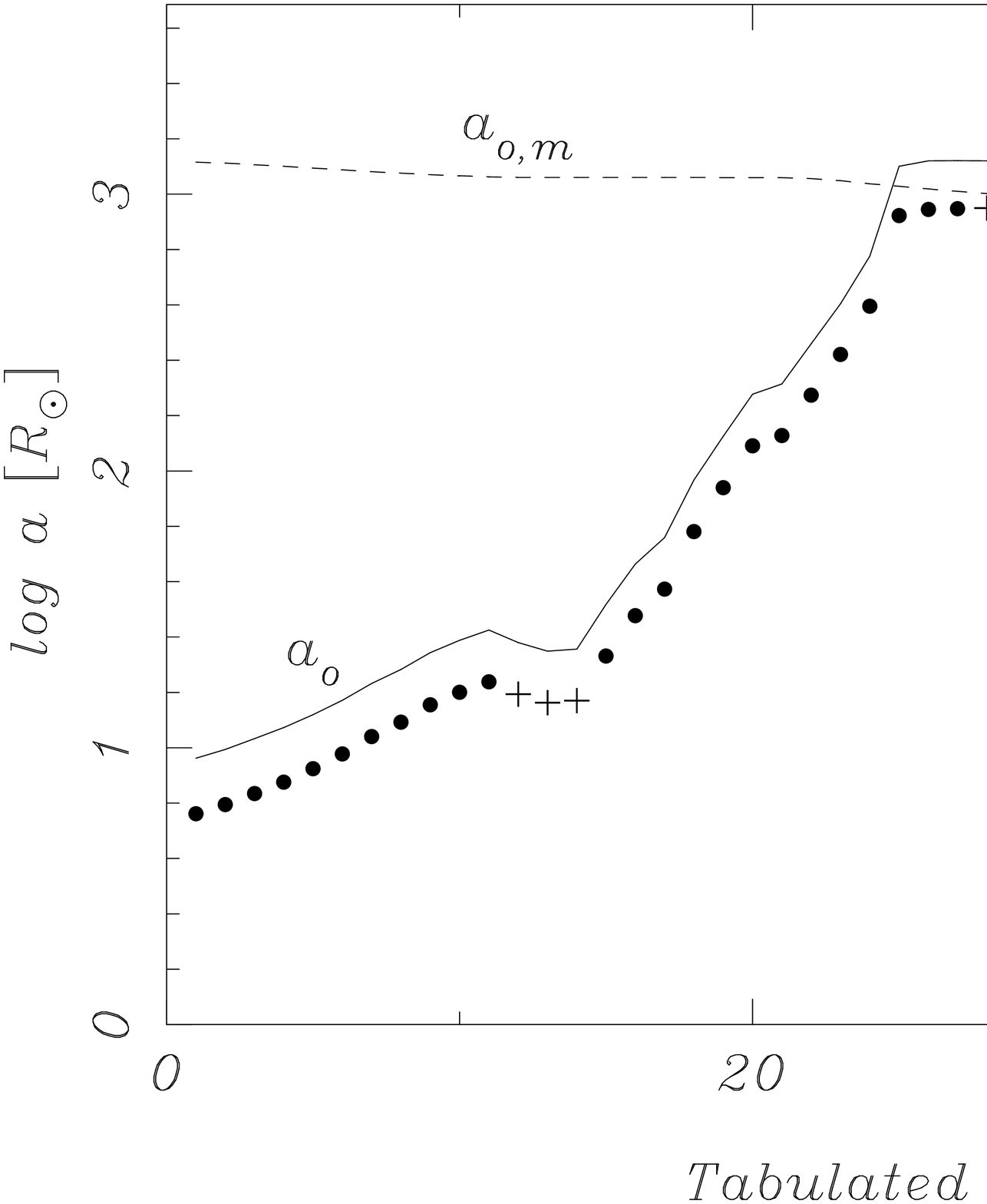} \\

\vspace*{0.5cm}

\hspace*{0.5cm}
\epsfxsize = 4.0cm
\epsffile{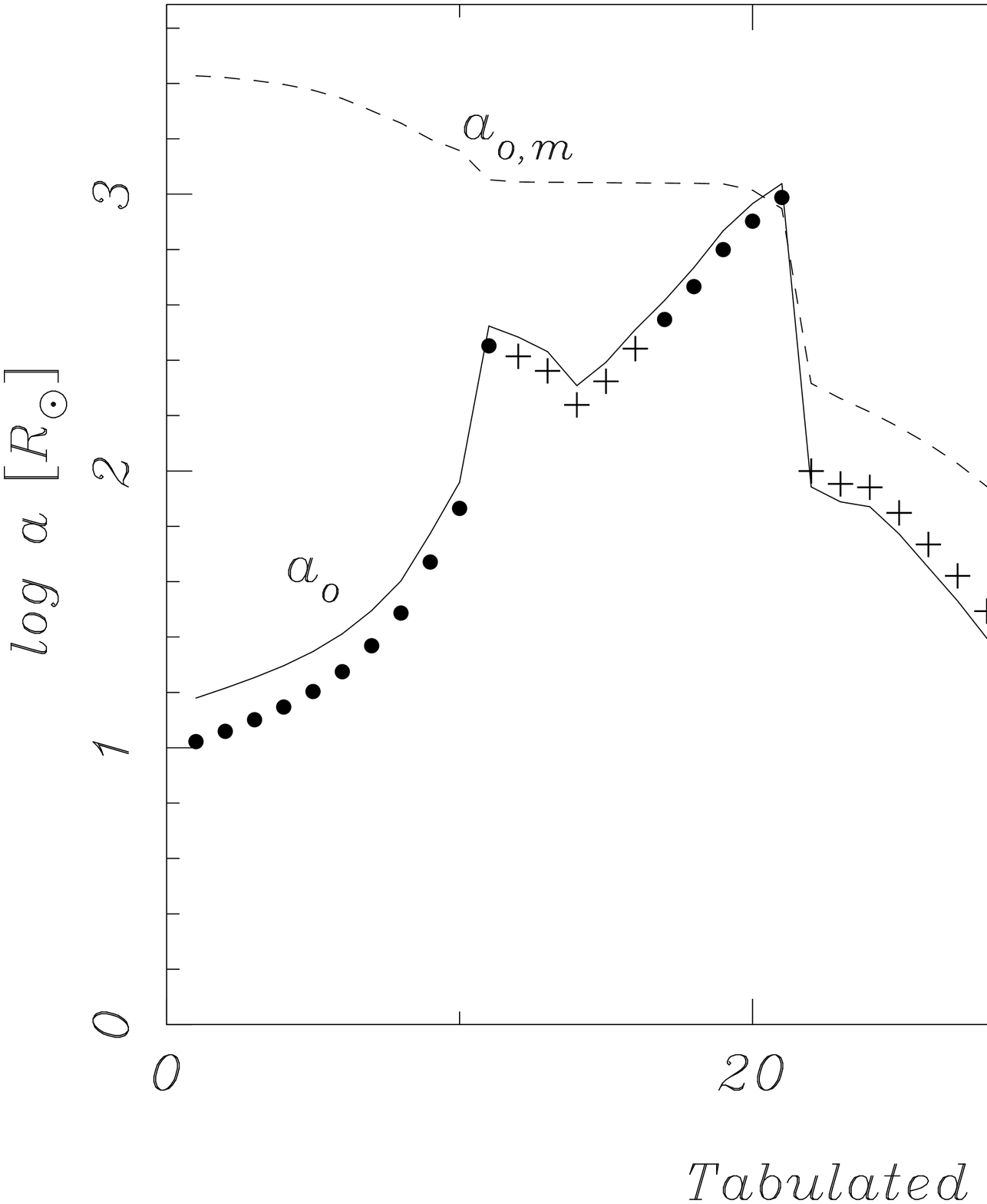}
\caption[]{
A: $\bullet$ and $+$ indicate radii for subsequent evolutionary 
stages of the 20~\msun\ primary as tabulated in \cite*{sch92}. 
From the mass of the primary at the tabulated point one may calculate
the semi-major axis of a binary with a
1~\msun\ secondary in which the primary fills its Roche lobe, 
and from this the semi-major axis $a_o$, shown as a solid line,
of the binary at the beginning of its evolution.
For each tabulated evolutionary stage 
we calculate the minimum semi-major axis at which the core of the
primary survives the spiral-in of a 1~\msun\ companion, and from this the
minimum semi-major axis $a_{o, m}$, indicated with the dashed line, 
of the binary at the beginning of its evolution.
Primaries at the evolutionary stages marked with a $+$ cannot fill their
Roche lobe for the first time at that stage but will reach their Roche
lobe at an earlier point in their evolution. 
A low mass X-ray binary is formed when $a_o>a_{o,m}$ for evolutionary
stages indicated with $\bullet$.

B: as A for a 60~\msun primary
} \label{Fig_lainitR}\end{figure}

Since the stellar parameters used in the previous Sect. are rather
rough, the same computation is performed with a more detailed model 
for population I ($Z=0.02$) stars.
We use the models with moderate core overshooting 
computed by \cite*{sch92}, using
the radius (calculated from effective temperature and luminosity) 
and the mass of the star in the tabulated points.
The mass of the star decreases as a function of time due to stellar
wind.
The mass loss in the stellar wind causes an increase of the
Roche-lobe radius of the primary, (mainly) by increasing the 
semi-major axis and (to a lesser extent) by increasing the mass-ratio
$m/M$. The increase in the semi-major axis is described, assuming an
isotropic wind with high velocity according to the Jeans approximation, 
with (\cite{vdh83}):
\begin{equation}
        a(M+m) = {\rm constant}.
\label{wind}\end{equation}
The relation between the semi-major axis $a_i$ at which the primary fills
its Roche lobe $R_L$ and the initial semi-major axis $a_o$ is thus given by:
\begin{equation}
        R_L = {\cal R}_L(q)a_i
            = {\cal R}_L\left(\frac{m}{M}\right) 
                            a_o\frac{M_o+m}{M+m}.
\label{semi_wind}\end{equation}
Here $M_o$ and $M$ are the zero-age mass of the primary and its mass at
the moment it fills its Roche lobe.
For each tabulated point of the evolutionary track we equate the radius
$R$ of the primary to the Roche-lobe radius $R_L$ in Eq.~\ref{semi_wind}
to calculate the corresponding maximum initial semi-major axis $a_o$.
The values for $a_o$ and $R$ for a 20 and a 60~\msun star accompanied
by a 1~\msun companion are shown in Fig.~\ref{Fig_lainitR}.
A value for $a_o$ smaller than reached at an earlier stage
of the evolution implies that Roche lobe overflow would have occurred 
at that earlier moment.

For each tabulated point of the evolutionary track we 
calculate the mass of the envelope 
$M_e$ by subtracting the core mass from the total mass.
Because the stellar evolution models incorporate overshooting, which
tends to increase the core mass, we calculate the core mass by
multiplying the value found from Eq.~\ref{Mcore} with 1.125 (\cite{mm89}).
We also know the minimum separation after spiral-in for a detached
binary.
From this we calculate the minimum separation at the
onset of spiral-in with Eq.~\ref{spiral_in}, and the minimum separation
of the initial binary $a_{o,m}$ with Eq.~\ref{semi_wind}.
These minimum separations are also shown in Fig.~\ref{Fig_lainitR}.

The value of $a_o$ of the first evolutionary point at which $a_o>a_{o,m}$, 
which we denote as $a_{min}$, corresponds to the minimum initial
separation that the binary must have to survive the spiral-in.
The maximum of the values of $a_o$ for all evolutionary stages, 
indicated with $a_{o,max}$, corresponds to the maximum initial 
separation of the binary at which the primary can reach its Roche lobe.
Only binaries with an initial separation in the range 
$a_{min}$ -- $a_{o,max}$ can evolve into low-mass X-ray binaries.
For a 20~\msun primary with a 1~\msun secondary this range is 
1000 -- 1590 \rsun. 

The Roche lobe can only be reached for the first time in 
those evolutionary stages for which $a_o$ is larger than
the $a_o$'s at all earlier evolutionary stages. 
Those stages are marked in Fig.~\ref{Fig_lainitR} with $\bullet$.
Note that core hydrogen burning ends in tabulated point 13, and 
helium core burning begins in point 21.
Helium core burning ends in point 43, and carbon burning starts in
point 46.
For a star of $\sim 5\msun$, the radius of the star expands
following the end of core hydrogen burning, and mass transfer during this
first ascent of the giant branch is called case~B. At the onset
of core helium burning, the star shrinks. It expands once more
after the end of core helium burning, and mass transfer during this
second ascent of the giant branch is called case~C.
For the $20\msun$ star shown in Fig.~\ref{Fig_lainitR},
however, the radius does not shrink at the onset of each new
phase of core fusion, but continues its expansion throughout its
evolution, once the Hertzsprung gap is passed.
The $60\msun$ shrinks at the onset of helium fusion in the core,
mainly due to extensive mass loss.

As shown by Fig.~\ref{Fig_lainitR} a 60~\msun\ primary with a 
1~\msun\ secondary can, according to the same reasoning, only 
evolve into a low-mass X-ray binary if its semi-major axis is in the
very small range of 980 -- 1100 \rsun.

\begin{figure}
\hspace*{0.5cm}
\epsfxsize = 4.0cm
\epsffile{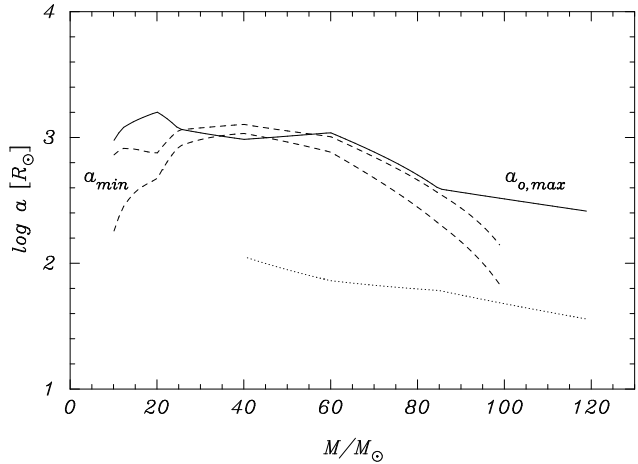}
\caption[]{Lower limit to the initial semi-major axis at which the
binary survives the spiral-in, as function of the initial mass of the
primary, calculated with use of the evolutionary sequences by
Schaller et al. (1992).
The upper (lower) dashed line gives the limit determined from the
condition that the secondary star (helium core) is smaller than its 
Roche-lobe, 
The solid line gives  the upper limit at which the primary 
reaches its Roche lobe at its maximum radius of 
$R_{max} = 1000\rsun$. The secondary is assumed to be a 1~\msun star.
The dotted line indicates the initial semi-major 
axis of a binary in which a 1~\msun\ secondary can fill its Roche lobe 
after the primary has lost its
entire envelope without ever having filled its Roche lobe}
\label{ce_survival}\end{figure}

We determine the mass
and size of primaries in a range of masses at the moment they fill their 
Roche lobes at $a_{o,max}$ for each tabulated stellar evolution track.
The masses and radii at $a_{o,max}$ of the stars that are not 
tabulated by \cite*{sch92} are determined by a linear interpolation
between the tabulated models. The resulting values for $a_{o,max}$
are shown as a solid line in Fig.~\ref{ce_survival}.

If Roche-lobe overflow for all initial semi-major axes smaller than 
$a_{o,max}$ leads to a merger, then $a_{min}$ is not properly defined.
A lower limit can be obtained by computing $a_{min}$
with the stellar parameters that correspond to the point where 
$a_{o,max}$ is reached.

Mass loss from stars more massive than $\sim 100\msun$ is so copious that
these stars lose their entire hydrogen envelope before they expand
on the giant branch, i.e.\ before Roche-lobe contact is achieved.
The common-envelope phase hardly leads to a spiral-in, whereas
further attrition of the core to a small final mass 
($\sim 8$~\msun\ according to \cite{sch92})
causes the binary orbit to expand.
As a result the final orbit is so wide that the 1~\msun\ secondary never
reaches its Roche lobe.

Figure~\ref{ce_survival} shows, as a function of
the zero-age mass of the primary, the lower limits
to the semi-major axis of the initial binary
at which the binary survives the spiral-in $a_{min}$ 
(or its conservative lower limit), and the upper limit
for which the primary reaches its Roche lobe $a_{o, max}$.

Comparison of Fig.~\ref{ce_survival} with Fig.~\ref{Fig_rom92}
illustrates that the mass loss of massive stars 
and the concurrent widening of the binary 
reduces the maximum initial separation for 
which the binary reaches Roche-lobe contact.
For stars with $\sim 70~\msun$\ the maximum stellar radius is smaller
than 1000~\rsun, which leads to a further reduction of $a_{o, max}$.
The wind mass-loss affects the minimum initial separation necessary
to survive the spiral-in in two ways: 
the widening of the orbit reduces this separation, whereas the reduction
of the envelope mass enlarges it.

Figure~\ref{ce_survival} indicates that the formation rate of
low-mass X-ray binaries with a neutron star greatly exceeds the
formation-rate of low-mass X-ray binaries with a black hole, because
the range of allowed initial separations is larger for neutron star
progenitors.

\cite*{rom92} mentions the possibility that
the secondary star might fill its Roche lobe as it evolves on the
asymptotic giant-branch after the primary collapsed into a remnant
without ever having reached Roche-lobe contact.
If this happens the binary orbit also widens dramatically, according to
Eq.~\ref{wind}. If the orbit widens too much, the secondary will
not reach its Roche lobe.
In Fig.~\ref{ce_survival} we show the maximum initial semi-major axis for which
a 1~\msun\ star can reach its Roche lobe after its companion has lost
its envelope.
It is seen that this semi-major axis is so small, that it invalidates
the assumption that Roche lobe contact of the primary has been avoided.

\section{Galactic formation rates}
The formation rate of binaries that reach
Roche-lobe contact and survive the spiral-in
can be estimated by computing the birth rate of binaries with 
suitable initial parameters; primary mass, mass ratio and semi-major axis.
For simplicity we neglect the eccentricity.
For the initial mass function for the primary 
we use a Salpeter function (\cite{sal64}) integrated over the
Galaxy:
\begin{equation}
\Psi_G(M) = 0.05 M^{-2.35} \, [\msun~{\rm yr}^{-1}].
\end{equation}
This normalization gives a supernova rate in agreement
with the observed rate.
The initial semi-major axis distribution
$\Gamma(a)$ is taken flat in $\log a$ (\cite{kpt+78}).
We assume a flat initial mass-ratio distribution, $\Phi(q)=1$.
For a given mass of the primary $M$ the fraction of binaries with a mass
of the secondary between $(m-\epsilon)$
and $(m+\epsilon)$ is then proportional to $2\epsilon/M$.

Primaries with an initial mass between 10 and 40~\msun\ are
assumed to leave a neutron star after the supernova, 
progenitors with a mass between 40 and 100~\msun\ form black holes.
The mass of the secondary is taken to be $1~\msun$.
The minima and maxima for the initial semi-major axis are
computed as described in the previous section.
We integrate the initial distribution functions for the primary mass, the
mass ratio and the semi-major axis between the mass 
limits $M_{min}$ and $M_{max}$ 
and the corresponding limits for the semi-major axis 
$a_{min}$ and $a_{o, max}$ (see Fig.~\ref{ce_survival}):
\begin{equation}
   Br = \int_{M_{min}}^{M_{max}} \int_{a_{min}}^{a_{o,max}}
        \Psi_G(M) \frac{2\epsilon}{M} \Gamma(a) \, da dM \, [{\rm yr}^{-1}].
\label{birthrate}\end{equation}
With $\epsilon = 0.15~\msun$,
with a binary fraction of 50\% of high mass stars,
and with the lower limit to $a_{min}$ set by the main-sequence star,
the galactic formation rate of binaries that reach Roche-lobe
contact and survive the spiral-in is
$2.2\cdot10^{-6}\, {\rm yr}^{-1}$ for the binaries with a neutron
star and $9.6\cdot10^{-9}\, {\rm yr}^{-1}$ for the
black-hole binaries. 
The formation rate thus calculated for the low-mass X-ray binaries
with a neutron star is compatible (within the rather wide
uncertainties) with the rate derived from the observed numbers of
such binaries.
The formation rate of low-mass X-ray binaries with a black hole is
about 1\%\ of the formation rate of low-mass X-ray binaries with a
neutron star, whereas the observations indicate equal formation rates
for these two types of binaries.


\section{Discussion}
We briefly discuss the assumptions and model uncertainties 
that affect our computations most.

In our calculations
the effect of convective overshooting on the mass of the helium core 
is taken into account in a relatively simple manner, since
the tabulated stellar evolution tracks do not provide masses
for the stellar helium core.
The importance of convective overshooting is uncertain, and therefore we
investigate small variations in its effects.

If we increase the mass of the core by increasing the multiplication factor
for Eq.~\ref{Mcore} discussed in paragraph 2 of Sect. 3 
from 1.125 to 1.250, the lower limit $a_{min}$ to the semi-major
axis for which a binary survives the spiral-in is slightly reduced,
and consequently the birthrate of neutron star binaries is enhanced
by $\sim 28\%$, and the birthrate of black hole binaries
increases with a factor $\la 4$.
An increase of the exponent in Eq.~\ref{Mcore} to 1.57,
which is not unrealistic according to \cite*{it85},
reduces the discrepancy between the 
birthrates for black holes and neutron stars in low-mass X-ray binaries
to a factor 15. Thus neither of these changes in Eq.~\ref{Mcore} leads to
a sufficiently high birthrate of black-hole binaries relative to 
neutron-star binaries.

\begin{figure}
\hspace*{0.5cm}
\epsfxsize = 4.0cm
\epsffile{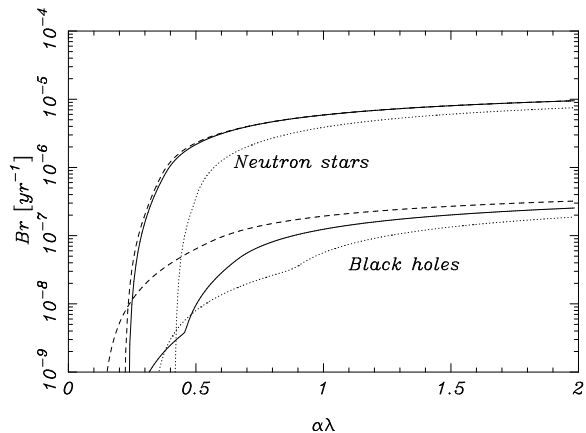}
\caption[]{Birthrate (per year in the galaxy)
of low-mass X-ray binaries with a neutron star
(upper lines) and a black hole companion (lower lines) as function of
the common-envelope efficiency-parameter $\alpha\lambda$.
The full lines are from the stellar evolution models from \cite*{sch92},
the dotted lines are from the low metallicity models with
$Z=0.004$ from \cite*{cha93} and the dashed lines are computed
with $Z=0.040$ from \cite*{sch93}}
\label{lmxb_br}\end{figure}

The efficiency $\alpha\lambda$ at which the common-envelope is expelled 
upon the spiral-in is highly uncertain. 
The effect of varying $\alpha\lambda$ on the galactic formation rate for
binaries that reach Roche-lobe contact
and survive the spiral-in is demonstrated in Fig.~\ref{lmxb_br}.
Decreasing $\alpha\lambda$ results, as expected,
in a strong reduction in the formation rate of low-mass X-ray
binaries with a neutron star as well as with a black hole.
By increasing $\alpha\lambda$ to a value larger than about unity, 
the formation rate of low-mass X-ray 
binaries approaches a constant rate, 
limited by the number of primordial binaries formed in the galaxy.
Note that $\alpha\lambda$ might have different
values for neutron star-- and black-hole progenitors.

To investigate the effect of the stellar metallicity on the
formation rate of low-mass X-ray binaries we compute two more models using
the evolutionary tracks calculated by \cite*{cha93} for $Z=0.004$ 
and \cite*{sch93} for $Z=0.04$ (see Fig.~\ref{lmxb_br}).
Apart from the enhanced mass-loss rates in the model with high metallicity 
which results in a larger formation rate of low-mass X-ray binaries, 
these model variations have little influence on our conclusions.
The models with a lower metallicity lose less mass in the stellar wind and
the fraction of binaries that survive the spiral-in decreases accordingly. 

The limit of $\sim 40 \msun$, which we use throughout the calculations, 
was originally derived from the observation of the high-mass
X-ray binary LMC X-3 (\cite{hh84}). 
Recently the analysis of the data of Wray~977, 
the BIa hypergiant companion of the X-ray pulsar
GX301-2, has increased the empirical lower mass limit for the 
formation of a black-hole in a binary to $M_o \sim 50~\msun$
(\cite{klr+95}).
This only exacerbates the problem of too low a formation rate of
black-hole binaries in our calculations.
\cite*{mae92} argues on the basis of the galactic ratio of the
abundances of metals and of helium that
the lower limit for the formation of a black
hole is 20 -- 25~\msun. As can be seen in Fig.~\ref{ce_survival}, such
decrease of the lower mass-limit has little effect on our computations,
because $a_{min}>a_{o, max}$ in the range of 25--40~\msun.

The stellar evolutionary tracks of \cite*{sch92} follow the single 
stars beyond the point of carbon burning. 
The large amount of mass and angular-momentum lost
in the stellar wind during earlier phases of the evolution of the 
more massive primaries cause the binary orbit to widen and prohibits 
mass transfer afte the onset of core helium burning
(see Fig.~\ref{Fig_lainitR}).
For the amount of angular momentum lost per unit mass in the stellar wind
we used the standard description of isotropic mass 
loss from the donor (see Eq.~\ref{wind}). 
Mass transfer at an evolutionary stage well after the onset of 
helium fusion (say, beyond
point 25 in Fig.~\ref{Fig_lainitR})
can only be an effective channel for the formation of low-mass 
X-ray binaries with a neutron star as the compact object, if the loss
of angular momentum is high enough for the orbit to shrink. 
For a more massive progenitor, larger than $\sim 40\msun$, which 
leaves a black hole instead of a neutron star after the supernova,
such late mass transfer
is not likely to become important even for such high loss of
angular momentum: the star has evolved into a naked helium core
and a merger will be unavoidable upon Roche-lobe contact.

The evolution of the helium star and the effect of the supernova on the 
binary orbit are neglected in our calculations:
both effects tend to lower the formation rate of low-mass X-ray
binaries (see e.g. \cite{pzv96}).
Mass transfer from the helium star to its companion can result in a
merger and the asymmetry of a supernova can dissociate the binary.
Both effects, however, tend to reduce the formation-rate of neutron stars
as well as black holes in low-mass X-ray binaries and do consequently 
not solve the birth-rate problem discussed in this paper.

\section{Conclusions}
Our computations, in contradiction to the results of \cite*{rom92},
predict a formation rate for low mass X-ray binaries
with a black hole which is
much smaller than the value derived from the observed numbers and
estimated X-ray lifetime.
The striking difference between our and his results can be seen
immediately by comparing his Fig.~1 with our Fig.~\ref{Fig_rom92}.  
Romani's values for $a_{max}$ are much higher than ours:
this is due to the fact that he uses $1/q$ instead of $q$ in
the equation for the Roche lobe (our Eq.~\ref{Roche_lobe}).
(This error has been silently corrected in \cite{rom94}.)
This is true both for binaries in which mass transfer preceeds the
supernova explosion, and for binaries in which the primary 
explodes after losing its envelope without ever having filled its 
Roche-lobe.
The discrepancy between theoretical and observed formation rates
cannot be solved by invoking different metallicities for the
progenitor systems, nor by assuming different efficiencies
for the envelope ejection during spiral-in.

Black-hole binaries can be produced in larger numbers only if
it is assumed that stars with initial masses less than approximately
20~\msun\ can collapse to black holes; or alternatively if it
is assumed that the angular momentum loss caused by the
stellar wind is so high that the binary orbit shrinks; or
alternatively if the collapse of a helium core to a black hole is
asymmetric, so that the post-supernova orbit can be smaller than
the pre-supernova orbit (see \cite{pzv96}).

\acknowledgements{This work was supported in part by the Netherlands
Organization for Scientific Research (NWO) under grant PGS 78-277.}

\bibliographystyle{aabib}
\bibliography{lmxb}

\begin{thebibliography}{}

\bibitem[\protect\astroncite{Charbonnel et~al.}{1993}]{cha93}
Charbonnel, C., Meynet, G., Maeder, A.~S., Schaller, G., Schaerer, D. 1993,
  A\&AS, 101, 415

\bibitem[\protect\astroncite{Cowley}{1992}]{cow92}
Cowley, A. 1992, ARA\&A, 30, 287

\bibitem[\protect\astroncite{de~Loore \& Doom}{1992}]{dld92}
de~Loore, C. W.~H., Doom, C. 1992,
\newblock Structure and evolution of single stars and binaries,
\newblock Kluwer Academic Publishers, Dordrecht, 179 edition

\bibitem[\protect\astroncite{Eggleton}{1983}]{egg83}
Eggleton, P. 1983, ApJ, 268, 368

\bibitem[\protect\astroncite{Iben \& Tutukov}{1985}]{it85}
Iben, I.~J., Tutukov, A.~V. 1985, ApJS, 58, 661

\bibitem[\protect\astroncite{Kaper et~al.}{1995}]{klr+95}
Kaper, L., Lamers, H., Ruijmakers, E., van~den Heuvel, E., Zuiderwijk, E. 1995,
  A\&A, 300, 446

\bibitem[\protect\astroncite{Kraicheva et~al.}{1978}]{kpt+78}
Kraicheva, Z.~T., Popova, E.~I., Tutukov, A.~V., Yungelson, L.~R. 1978, AZh.,
  55, 1176

\bibitem[\protect\astroncite{Maeder}{1992}]{mae92}
Maeder, A. 1992, A\&A, 264, 105

\bibitem[\protect\astroncite{Maeder \& Meynet}{1989}]{mm89}
Maeder, A., Meynet, G. 1989, A\&A, 210, 155

\bibitem[\protect\astroncite{Portegies~Zwart \& Verbunt}{1996}]{pzv96}
Portegies~Zwart, S.~F., Verbunt, F. 1996, A\&A, 309, 179

\bibitem[\protect\astroncite{Romani}{1992}]{rom92}
Romani, R.~W. 1992, ApJ, 399, 621

\bibitem[\protect\astroncite{Romani}{1994}]{rom94}
Romani, R.~W. 1994,
\newblock in A.~W. Shafter (ed.), Interacting Binary Stars, ASP Conf.\ Ser.\
  56, p.~196

\bibitem[\protect\astroncite{Salpeter}{1964}]{sal64}
Salpeter, E.~E. 1964, ApJ, 140, 796

\bibitem[\protect\astroncite{Schaerer et~al.}{1993}]{sch93}
Schaerer, D., Charbonnel, C., Meynet, G., Maeder, A.~S., Schaller, G. 1993,
  A\&AS, 102, 339

\bibitem[\protect\astroncite{Schaller et~al.}{1992}]{sch92}
Schaller, G., Schaerer, D., Meynet, G., Maeder, A.~S. 1992, A\&AS, 96, 269

\bibitem[\protect\astroncite{Tanaka \& Lewin}{1995}]{tl95}
Tanaka, Y., Lewin, W. H.~G. 1995,
\newblock in W.~H. G. e.~a. Lewin (ed.), X-ray binaries, Cambridge University
  Press,  536

\bibitem[\protect\astroncite{Tanaka \& Shibazaki}{1996}]{ts96}
Tanaka, Y., Shibazaki, N. 1996, ARA\&A, 34, in press

\bibitem[\protect\astroncite{van~den Heuvel}{1983}]{vdh83}
van~den Heuvel, E. 1983,
\newblock in W. Lewin, E. van~den Heuvel (eds.), Accretion-driven stellar
  {X}-ray sources, Cambridge U.P., Cambridge, p.~303

\bibitem[\protect\astroncite{van~den Heuvel \& Habets}{1984}]{hh84}
van~den Heuvel, E., Habets, G. 1984, Nat, 309, 598

\bibitem[\protect\astroncite{Webbink}{1984}]{web84}
Webbink, R.~F. 1984, ApJ, 277, 355

\end{thebibliography}
\end{document}